\documentclass{raa}

\usepackage{graphicx,times}             

\begin{document}

   \title{On the emission lines in active galactic nuclei with relativistic jets}

   \volnopage{Vol.0 (200x) No.0, 000--000}      
   \setcounter{page}{1}          

   \author{L. Foschini}

   \institute{INAF -- Osservatorio Astronomico di Brera, Via E. Bianchi 46, 23807, Merate (LC), Italy; {\it luigi.foschini@brera.inaf.it}}

   \date{Received~~2011 October 7; accepted~~year~~month day}

\abstract{The effect of the observed continuum emitted from a relativistic jet on the measurement of the full width half maximum (FWHM) of an emission line is analyzed. If the jet contribution is not properly subtracted, the FWHM of the line could seem narrower than what it should be. The cases of emission line detected in BL Lac objects and $\gamma$-ray Narrow-Line Seyfert 1 galaxies ($\gamma$-NLS1s) are addressed. It is shown that the smallness of the observed FWHM of the Ly$\alpha$ lines detected in three well-known BL Lacs, is an effect due to the combined action of both the relativistic jet and a weak accretion disc. Once removed the Doppler boosting of the jet continuum, the intrinsic FWHM of the lines are found to be in the usual range. Instead, the narrow permitted lines in $\gamma$-NLS1s are really narrow, since the disc and the lines are much more powerful. This also confirms that $\gamma$-NLS1 is really a new class of $\gamma$-ray emitting AGN, different from blazars and radio galaxies.
}

   \authorrunning{L. Foschini}            
   \titlerunning{Emission lines in AGN with jets}  

   \maketitle

\section{Introduction}
The recent detection by Stocke et al. (2011) of weak (equivalent width EW~$<1$~\AA) and narrow (FWHM $\sim 300-1000$~km/s) Ly$\alpha$ lines from three TeV BL Lac Objects (Mkn 421, Mkn 501, PKS~2005$-$489) poses important questions on their spatial origin and how they are generated. Indeed, the FWHM is not only a measurement of the kinetic conditions of the plasma nearby the central singularity, but it is also an estimator of the black hole mass mainly through the reverberation mapping technique (Blandford \& McKee 1982, Peterson et al. 1998, Wandel et al. 1999 and many more). Given the density and temperature conditions inferred from emission lines measurements, the broadening due to thermal energy and turbulence has not a significant impact (e.g. Netzer 1990; see however Foschini 2002). Therefore, the FWHM is mostly dependent on the bulk motion of the plasma and can be used to calculate the mass $M$ of the black hole under the virial assumption: 

\begin{equation}
M = \frac{R \cdot f\cdot v_{\rm FWHM}^2}{G}
\label{virial}
\end{equation}

\noindent where $f$ is an unknown parameter linking the FWHM to the plasma bulk motion speed in the broad-line region (BLR, $v_{\rm BLR}=\sqrt{f}\cdot v_{\rm FWHM}$), $R$ is the radius of the BLR and $G$ is the Newton's gravitational constant. In the most general case of a Keplerian motion in a spherically BLR, $f=3/4$ (Netzer 1990), while a disc-like BLR viewed with an angle $\Theta$ has $f=(4\sin^2 \Theta)^{-1}$ (McLure \& Dunlop 2002). Other values have been proposed, by taking into account systematic effects (e.g. Collin et al. 2006).

\begin{figure*}
\begin{center}
\includegraphics[angle=270,scale=0.35]{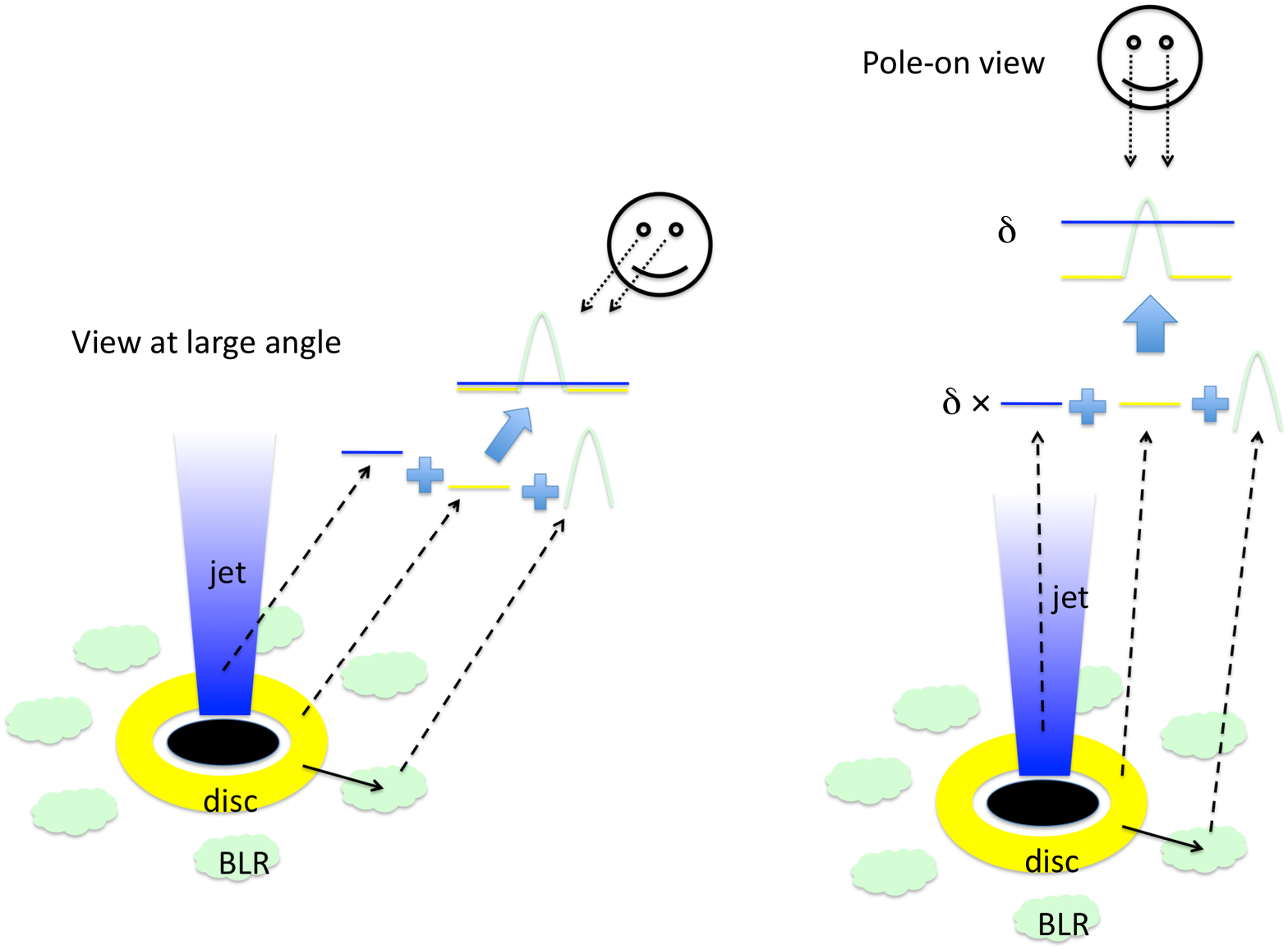} 
\caption{Sketch (not in scale) of the emission components of an AGN with relativistic jet. (\emph{left}) View from large angles: the relativistic jet gives small contribution to the observed continuum, because the Doppler boosting is low. (\emph{right}) View from small angles: the effect of special relativity boosts the intrinsic continuum of the jet, which is now no more negligible and can overwhelm the emission line, making it very difficult to measure its characteristics. If the observed jet contribution is not properly removed, then the FWHM of the line seems to be narrower than its real value.}
\label{fig1}
\end{center}
\end{figure*}

By means of statistical studies, several authors have found that broad lines emitted from AGN with relativistic jets have FWHM a bit smaller than those from AGN with no jets (Wills \& Browne 1986, Vestergaard et al. 2000, Jarvis \& McLure 2006, Fine et al. 2011). Specifically, quasars whose radio emission is dominated by the core have the narrower lines and this is explained as an orientation effect and geometric shape of the BLR. Taking into account that core-dominated quasars (FSRQ) are viewed pole-on, while lobe-dominated quasars are viewed at much larger angles, if the BLR has an equatorial disc-like shape, then the FWHM is smaller than in the case of a spherical BLR or if the source is viewed edge-on, because the kinetic component directed to the observer is missing or negligible.

In addition, the Doppler boosting of the intrinsic jet continuum can play an important role (see Fig.~\ref{fig1}). It is reasonable to expect that there could be cases where the observed jet emission can overwhelm the line emission (blazars: pole-on view $\rightarrow$ small viewing angles $\rightarrow$ Doppler boosting), which in turn \emph{seemingly} changes its FWHM. It is known that the functions adopted to fit the line emission profiles are self-similar (e.g.  gaussian), but if the continuum level is too high -- and thus the EW is very small -- then the measurement of the FWHM could really give misleading results if the observed jet contribution is not properly taken into account.

Exemplary is the case of BL Lac reported by Corbett et al. (2000): depending on the observed jet continuum emission, the measured FWHM of the H$\alpha$ line changes from 5050~km/s (1997 November 14) to 2030~km/s (1997 December 7) in about 24 days. The line disappeared at the maximum observed continuum flux on 1997 June 27 (see Table~2, Fig.~2 and 3 of Corbett et al. 2000). Although the jet and the line emission are correlated, they are not physically linked, as also stated by Corbett et al. themselves. It is possible to prove it by \emph{reductio ad absurdum}. Let us make the hypothesis that the measured changes in the FWHM are real and not due to measurement problems caused by the high observed continuum. This means that the bulk motion of the BLR is changed. By assuming a mass of the central black hole of BL Lac as $5\times 10^{8}M_{\odot}$ (Ghisellini et al. 2010) and by adopting the virial factor $f$ of McLure \& Dunlop (2002) calculated with an angle $\Theta = 3^{\circ}$, then it is possible to estimate the radius of the BLR as $2.7\times 10^{15}$~cm in the case of FWHM~$= 5050$~km/s and $1.7\times 10^{16}$~cm in the case of FWHM~$= 2030$~km/s. This would mean that the BLR size changed of about one order of magnitude in less than one month, which is not reliable, because it requires that the plasma of the whole BLR should have had an outward radial motion of the order of $0.2c$. Such a massive outflow needs of a much stronger disc than that available in BL Lac to be produced (cf Pounds \& Page 2004).  Therefore, the simplest possibility (\emph{lex parsimoniae}) is to think that the observed changes are not real, but due to the superimposed boosted continuum from the relativistic jet, which was not properly taken into account. It is worth underlining that the key issue is the time scale: indeed, significant changes of the FWHM have been observed in AGN without jets and are thought to be due to real dynamical changes (e.g. Wanders \& Peterson 1996). However, in the latter case, the time scale is of the order of years and not a few tens of days as for BL Lac.

The main aim of this work is to find a suitable correction to apply to the FWHM when it is not possible to properly subtract the observed jet continuum, because of the weakness of the line. It is shown that when the observed jet contribution is removed, then the real FWHM of these lines is in the usual range expected from broad permitted lines and is no more correlated with the increase or decrease of the observed jet emission. Therefore, their narrowness is just an observational effect due to the jet Doppler boosting. The basic concepts are explained in the Sect.~2, while specific application to the case of BL Lac Objects and $\gamma-$NLS1s are dealt in the Sect.~3 and 4, respectively. Some final remarks in Sect.~5 conclude the work.

\section{Basic concepts}
The BLR is ionized by the ultraviolet photons of the accretion disc and therefore the emission line flux variability is dependent on the accretion disc power (reverberation mapping, Blandford \& McKee 1982). The relativistic jet has negligible impact in generating lines, and hence in their variability, because its radiation is strongly beamed within a small angle ($\theta \sim \Gamma^{-1}$, where $\Gamma$ is the bulk Lorentz factor) on a direction almost perpendicular to the disc plane. Although the jet and the disc are believed to be somehow connected (e.g. Ghisellini \& Tavecchio 2008), their time scales are different, because physical mechanisms at work are different. Therefore, the emission lines flux variability is generally unrelated to the changes in the jet emission, as shown by the case of BL Lac (Corbett et al. 2000). This means that -- on short time scales -- even if the line holds its flux and profile, the observed continuum can increase because of the changes in the jet Doppler boosting (which in turn does not affect the line), thus reducing the line EW, but also seemingly its FWHM (Fig.~\ref{fig1}). 

It is important to stress this point: the line flux is strictly linked to the disc power (reverberation mapping, Blandford \& McKee 1982) and, therefore, a change in the disc continuum results in a change of the line flux. The intrinsic FWHM does not change on short time scales, as it is an indication of the keplerian motion of the BLR plasma (\footnote{Since the thermal broadening is negligible when compared to the bulk motion, a change in the disc luminosity has no impact on the line profile. Dynamical changes implying variations of the line profile occur on much longer time scales (e.g. Wanders \& Peterson 1996).}). When a relativistic jet is present, we are observing a Doppler-boosted continuum superimposed on the disc continuum (Fig.~\ref{fig1}). Its beamed flux can surpass the disc and line flux (that are not boosted), resulting in a \emph{seemingly change} of the observed FWHM of the line. It is worth noting once again that the intrinsic properties of the line remain stable on the jet time scale, because of their dependence on the disc power and the bulk motion of the BLR (see Fig.~\ref{fig1}).

To calculate the order of magnitude of the involved effects, let us first consider an AGN with a jet viewed at large angles, so that the effects of special relativity are negligible (Fig.~\ref{fig1}, \emph{left panel}). There is the accretion disc generating a ionizing continuum and the emission lines of the BLR, which can be described by a gaussian profile with width $\sigma$. It is well-known that the intrinsic FWHM is related to $\sigma$ by: 

\begin{equation}
{\rm FWHM_{\rm int}} = 2\sigma\sqrt{2\cdot(-\ln \frac{1}{2})}\sim 2.35\sigma
\label{eq:fwhm}
\end{equation}

Roughly speaking, the ratio between the line luminosity and the continuum is the EW:

\begin{equation}
{\rm EW} \sim L_{\rm line}/L_{\rm cont}
\label{blabla}
\end{equation}

\noindent where $L_{\rm cont}$ is integrated in a proper frequency range around the line frequency and is composed of a disc and a jet contribution ($L_{\rm cont}=L_{\rm disc}+L_{\rm jet}$). The two contributions could be at maximum of the same order of magnitude (Rawlings \& Saunders 1991; Ghisellini et al. 2010, particularly see Fig.~6).

When the jet viewing angle $\vartheta$ is small, then the special relativity significantly affects the radiative output (Fig.~\ref{fig1}, \emph{right panel}). The jet luminosity is amplified by a factor $\delta^4$ in the case of a spherical blob or $\delta^3$ for a steady jet, where $\delta=[\Gamma(1-\beta\cos\vartheta)]^{-1}$ is the Doppler factor and $\beta=v/c$. When compared to the accretion disc luminosity, the latter results to be overwhelmed. Such ``intrusive'' presence has the effect to alter the observed characteristics of the emission lines, because the observed continuum is now dominated by the boosted emission of the jet. Obviously, there could be different grades of overwhelming, basically depending on two different factors: the strength of the accretion disc and the frequency of the synchrotron peak. FSRQs have strong discs and low peak frequencies (optical/IR): therefore, the jet overwhelms the disc and the lines only during intense outbursts. BL Lac objects have weak discs and the synchrotron peaking at UV frequencies: therefore, the jet is always dominating over the disc and the lines.

Presently, it is important to focus on the basic concept: when the boosted jet continuum is superimposed, the observed FWHM could not match the intrinsic one. More generally, the observed FWHM will be equivalent to that measured at some fraction $q$ of the flux:

\begin{equation}
{\rm FWHM_{\rm obs}} = 2\sigma\sqrt{2\cdot(-\ln q)}
\label{eq:fwq}
\end{equation}

\noindent with $1/2<q<1$. The ratio between FWHM$_{\rm int}$ and FWHM$_{\rm obs}$ can be evaluated by imposing the equivalence of $\sigma$ (i.e. the line is the same and is costant) in Eqs.~(\ref{eq:fwhm}-\ref{eq:fwq}):

\begin{equation}
\frac{{\rm FWHM_{\rm int}}}{{\rm FWHM_{\rm obs}}} = \frac{\sqrt{\ln 2}}{\sqrt{-\ln q}}
\label{eq:ratio}
\end{equation}

It is necessary to find a suitable expression of $q$ that have to indicate the level of the observed jet contribution with respect to the line peak flux (fraction of the maximum flux). If one knows the Doppler factor $\delta$ from other methods (e.g. by modeling the spectral energy distribution -- SED), then it is sufficient to perform a correction of the continuum. Instead, in the present work, I would like to study the possibility to use the equivalent width EW of the line. 

Indeed, the boosted jet continuum alters also the EW: the intrinsic jet continuum is boosted by $\delta^4$, if the jet can be modeled as a spherical blob, while the disc luminosity -- not boosted -- is overwhelmed by the jet and therefore it can be neglected. The observed continuum is dominated by the jet: $L_{\rm cont} = L_{\rm disc} + \delta^4L_{\rm jet} \sim \delta^4L_{\rm jet}$. Compared to the unbeamed case above, it is evident that the ratio of the two continua is dominated by the Doppler factor:

\begin{equation}
\frac{{\rm EW_{\rm obs}}}{{\rm EW_{\rm exp}}} = \frac{L_{\rm line}}{(L_{\rm disc}+\delta^4L_{\rm jet})}\frac{(L_{\rm disc}+L_{\rm jet})}{L_{\rm line}}\sim \frac{1}{\delta^4}
\label{blabla2}
\end{equation}

\noindent and, consequently, the observed EW is reduced by a factor $\delta^{4}$. Therefore, it is possible to use the ratio between the observed EW (EW$_{\rm obs}$) and the expected value in the case of absent or negligible jet contribution (EW$_{\rm exp}$) to bypass the direct knowledge of $\delta$.

A few words on the factor $\delta^4$: as known, this is in the case of a spherical blob. In the case of a steady jet, the factor is $\delta^3$. This has no impact either on the above assumptions or on the calculation of the correction factor for the FWHM (see below), because it is just based on the ratio between the observed and expected EW; it has impact on the value of $\delta$ that can be calculated from this ratio. 

When inserting the EW ratio in $q$, it is necessary to take into account that $q=1/2$ if EW$_{\rm obs}=$EW$_{\rm exp}$ (i.e. FWHM$_{\rm int}=$FWHM$_{\rm obs}$) and $q\rightarrow 1$ when EW$_{\rm obs}<<$EW$_{\rm exp}$. One possible function satisfying the necessary constraints is (though other functions cannot be excluded):

\begin{equation}
q=1-\frac{{\rm EW}_{\rm obs}}{2{\rm EW}_{\rm exp}}
\label{parapoz}
\end{equation}

Therefore, by substituting Eq.~(\ref{parapoz}) in Eq.~(\ref{eq:ratio}), it results:

\begin{equation}
{\rm FWHM_{int}} = {\rm FWHM_{obs}}\frac{\sqrt{\ln 2}}{\sqrt{-\ln(1-\frac{{\rm EW}_{\rm obs}}{2{\rm EW}_{\rm exp}})}}
\label{fwhm}
\end{equation}

As displayed in Fig.~\ref{fig3}, when EW$_{\rm obs}\rightarrow$~EW$_{\rm exp}$ the correction is more and more negligible and unnecessary, since the direct measurement already gives a value consistent with the intrinsic one. The correction is roughly a factor 2 when EW$_{\rm obs}$ is about 30\% of EW$_{\rm exp}$. The smaller is the ratio, the larger is the correction. 

\begin{figure}[!t]
\begin{center}
\includegraphics[angle=270,scale=0.35]{ms948fig2.ps} 
\caption{FWHM$_{\rm int}$/FWHM$_{\rm obs}$ ratio as a function of the EW$_{\rm obs}$/EW$_{\rm exp}$ ratio.}
\label{fig3}
\end{center}
\end{figure}

\begin{figure}[!t]
\begin{center}
\includegraphics[angle=270,scale=0.35]{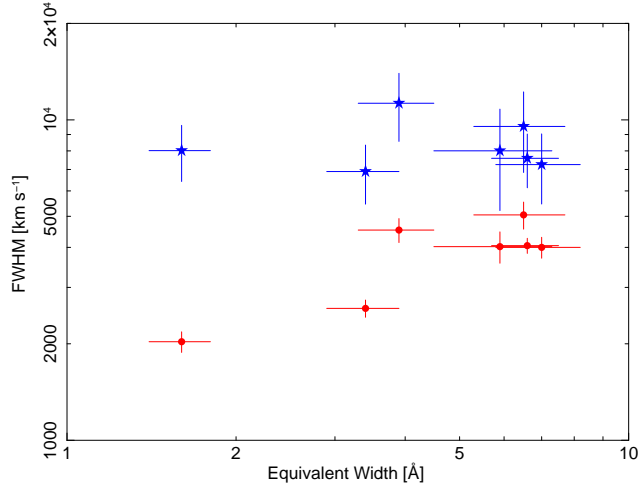} 
\caption{FWHM vs EW of BL Lac. The red circles are the measurements (FWHM$_{\rm obs}$) by Corbett et al. (2000), while the blue stars are the measurement corrected for the jet contribution (FWHM$_{\rm int}$). It is immediately evident that the red circles indicate a narrower FWHM$_{\rm obs}$ as the EW decreases (i.e. the observed jet contribution increases), while the corrected values (FWHM$_{\rm int}$) are no more dependent on the EW.}
\label{fig2}
\end{center}
\end{figure}

To make an immediate example, I apply this correction to the measurements of the H$\alpha$ line of BL Lac reported by Corbett et al. (2000). The authors themselves noted that the measured changes in the FWHM are biased by the high level of continuum due to the jet and the intrinsic profile of the line should be remained almost constant. However, they did not attempt any correction and the fit of the original FWHM$_{\rm obs}$ measured by Corbett et al. (2000) to a constant value of FWHM gives a reduced $\chi^2$ equal to 20.1 (i.e. not consistent with a constant value). To apply the Eq.~(\ref{fwhm}), I adopt EW$_{\rm exp}=$18~\AA, which refers to the H$\alpha$ broad component only (Constantin \& Shields 2003). It results immediately a \emph{broader} and \emph{constant} FWHM$_{\rm int}$ (Fig.~\ref{fig2}). The intrinsic FWHM, calculated as the weighted average of the corrected values, is $7857\pm 1208$ km/s and the reduced $\chi^2$ test for the variability of the FWHM$_{\rm int}$ vs EW gives now 0.43 (i.e. it is consistent with a constant value).

\begin{table}
\begin{center}
\caption{Summary of the observed quantities of the Lyman $\alpha$ line: $L_{\rm Ly\alpha}$ [$10^{41}$~erg~s$^{-1}$], EW$_{\rm obs}$ [\AA] and FWHM$_{\rm obs}$ [km~s$^{-1}$] from Stocke et al. (2011), with the corresponding calculated FWHM$_{\rm int}$ [km~s$^{-1}$] and Doppler factors for beaming, according to the procedure outlined in the present work and having considered EW$_{\rm exp}=40-100$~\AA. The size of the BLR is in units of [$10^{15}$~cm]. The mass $M$ is calculated with $f$ proposed by McLure \& Dunlop (2002) and assuming a viewing angle of $3^{\circ}$.}
\label{Tab:tabella1}
\begin{tabular}{lccccccc}
\hline
Source & $L_{\rm Ly\alpha}$ & EW$_{\rm obs}$ & FWHM$_{\rm obs}$ & FWHM$_{\rm int}$ & $\delta$ & $R_{\rm BLR}$ & $\log M/M_{\odot}$\\
\hline
Mkn~421 & 0.24 & 0.076 & 300 & 8122-12812 & 4.8-6.0 & 1.5 & 8.8-9.2\\
Mkn~501 & 0.52 & 0.830 & 820 & 6702-10586 & 2.6-3.3 & 2.3 & 8.8-9.2\\
2005$-$489 & 2.49 & 0.467 & 1050 & 11453-18080 & 3.0-3.8 & 5.0 & 9.6-10.0\\
\hline
\end{tabular}
\end{center}
\end{table}

\section{Application to BL Lac Objects}
Stocke et al. (2011) recently reported the detection of very weak (EW~$<1$~\AA) Ly$\alpha$ emission lines from three well-known TeV BL Lac Objects: Mkn~421 ($z=0.030$), PKS~2005$-$489 ($z=0.071$) and Mkn~501 ($z=0.0337$). Since Ly$\alpha$ are thought to be produced in the BLR, FWHM of several thousands of km/s are expected (masses of BL Lacs are $\sim 10^{8-9}M_{\odot}$). Instead, Stocke et al. (2011) found values for the Ly$\alpha$ FWHM of 300, 1050, and 820 km/s, for the three blazars respectively. They suggested different possible explanations based on the covering factor of the BLR, the ionization power, or that the emitting plasma is located far away from the black hole, at distances in excess of 10~pc (i.e. in the NLR). Another possible explanation is that if these FWHM are indeed small and produced in the BLR, this would imply small masses for these BL Lacs, with many important consequences. Or -- this is indeed what I am going to proof -- the lines are broad, with FWHM within the usual range, but they are observed to be narrow because of the presence of the jet emission. Something similar to the case of BL Lac analyzed in Sect.~2, although more extreme.

The values of the EW measured by Stocke et al. (2011) in the case of the three BL Lacs are of the order of tens/hundreds of m\AA, much lower than the average values of $40-100$~\AA, found in other types of AGN (Vanden Berk et al. 2001, Telfer et al. 2002, Constantin \& Shields 2003, Bachev et al. 2004, Pian et al. 2005, Gavignaud et al. 2006). Stocke et al. (2011) made the hypothesis that the ionizing radiation generating the Ly$\alpha$ is provided by the jet, because it is commonly though that the disc of BL Lac Objects is inefficient, an advection-dominated accretion flow (ADAF). However, although the disc could have low luminosity, it is likely \emph{not} in an advection-dominated regime yet. Elitzur \& Ho (2009) suggested that the BLR vanishes (which in turn occurs when the disc is radiatively inefficient) when the disc power is below a critical value:

\begin{equation}
L_{\rm crit} = 5\times 10^{39}(\frac{M}{10^{7}M_{\odot}})^{2/3}\, {\rm erg/s}
\end{equation}

By considering masses for Mkn~421, Mkn~501, and PKS~2005$-$489 with values of $4\times 10^{8}$, $10^{9}$, and $4\times 10^{8}$~$M_{\odot}$, respectively (from Wagner 2008), then the critical disc luminosities for the disappearance of the BLR are $6\times 10^{40}$, $10^{41}$, and $6\times 10^{40}$~erg/s, respectively. These values are comparable with the luminosities of the Ly$\alpha$ lines observed by Stocke et al. (2011), which are: $2.4\times 10^{40}$, $5.2\times 10^{40}$, and $2.5\times 10^{41}$~erg/s for Mkn~421, Mkn~501, and PKS~2005$-$489, respectively. Therefore, since the disc must have a greater luminosity -- say at least one order of magnitude greater taking into account the covering factor (e.g. Netzer 1990) -- to generate such lines, it results that the discs of these three BL Lac Objects are still powerful enough to be considered in the standard regime. 

These BL Lacs have lines and disc luminosities that are 3$-$4 orders of magnitude smaller than the average in powerful blazars ($<L_{\rm Ly\alpha}>=10^{44-45}$~erg/s, see Pian et al. 2005), but still similar to the latter, i.e. the disc is still standard and not dominated by the advection. It is in agreement with the findings of Maoz (2007) and Pian et al. (2010), according to which the low luminosity AGN have accretion discs similar to the more luminous Seyferts cousins and therefore replicates the same patterns although at low fluxes. This also support the use of the EW in calculating the real FWHM of BL Lacs: the disc and lines are the same as powerful blazars, but scaled down to low luminosities, although still in a standard disc regime. It is reasonable to think that the EW remains more or less the same along several orders of magnitudes of disc and lines luminosities. 

Therefore, from the ratio of EW$_{\rm obs}$/EW$_{\rm exp}$, it is possible to estimate the real FWHM$_{\rm int}$ and the Doppler factor $\delta$ necessary to increase the continuum in order to reach the observed EW, under the hypothesis that the line remains almost constant during the changes of the jet emission. The results are summarized in Table~\ref{Tab:tabella1}. The FWHM$_{\rm int}$ are now typical of lines from a BLR and the Doppler factors, although lower than the values obtained by Stocke et al. (2011), are more consistent with a relatively low activity of the jet during the observations. Indeed, no outbursts have been reported in the days of the measurements and, perhaps, this is the reason for which it was possible to detect these weak lines. The SEDs of these three blazars reported by Tavecchio et al. (2010) indicate that when the jet is active, its ultraviolet luminosity can even surpass the observed Ly$\alpha$ power, resulting in an observed featureless spectrum.

As for additional check, it is now possible to calculate the masses of the central black holes. The radius of the BLR can be estimated by the relationship (Ghisellini \& Tavecchio 2008):

\begin{equation}
R_{\rm BLR} = 10^{17}\sqrt{\frac{L_{\rm disc}}{10^{45}\, \rm erg/s}}\, \rm cm
\end{equation}

\noindent and by assuming a disc luminosity about one order of magnitude greater than the Ly$\alpha$ power (e.g. Netzer 1990). Given the low luminosity of the disc, the radius of the BLR is much smaller than in powerful blazars and, despite the correction on the FWHM, the mass calculated by Eq.~(\ref{virial}) and $f=1$, resulted in a value two orders of magnitudes smaller than the quantities measured with other methods (cf Wagner 2008). By considering a disc-like BLR and an almost pole-on orientation ($\Theta = 3^{\circ}$, see Ghisellini et al. 2010), the value of $f$ as proposed by McLure \& Dunlop (2002) results in a factor $f=(4\sin^2 \Theta)^{-1}\sim 90$. The masses are now consistent with the other values reported by Wagner (2008). The results are summarized in Table~\ref{Tab:tabella1}.

\begin{table}[!t]
\begin{center}
\caption{PMN~J0948$+$0022: comparison of flux densities [in units of $10^{-16}$~erg~cm$^{-2}$~s$^{-1}$~\AA$^{-1}$] as measured in 2000 by the SDSS (MJD 51602/51630) and in 2009 during the MW Campaign. $f_{\rm cont}$ indicates the continuum at $7705$~\AA \, as measured from the SDSS spectrum (average of the two observations). $f_{\rm I}$ and $f_{\rm R}$ are the flux densities in the $I$ and $R$, respectively. In the case of SDSS, the conversion from $ugriz$ to $UBVRI$ has been done according to Chonis \& Gaskell (2008). In the case of the 2009 MW Campaign the $f_I$ has been extrapolated from $B$ and $R$ fluxes. All the fluxes have been dereddened by using $N_{H}=5.22\times 10^{20}$~cm$^{-2}$ (Kalberla et al. 2005) and standard extinction laws (Cardelli et al. 1989).}
\begin{tabular}{lccc}
\hline
Period & $f_{\rm cont}$ & $f_{\rm I}$ & $f_{\rm R}$ \\
\hline
51602/51630   & 1.1 & 0.8 & 1.1\\
54956   & {}   & 2.9  & 4.0\\
54966   & {}   & 1.2  & 1.8 \\
\hline
\end{tabular}
\end{center}
\label{0948}
\end{table}

\section{The case of Narrow-Line Seyfert 1 Galaxies}
In the light of the considerations exposed above, one could ask oneself if the narrow permitted lines of $\gamma$-NLS1 (see Foschini 2011 for a review) are really narrow or if it is an effect of the presence of the jet, as in the case of BL Lacs above mentioned. The latter option is likely to be easily discarded, as the disc of $\gamma$-NLS1 is much more powerful than BL Lacs, similar to FSRQs, and hence the lines are more prominent and easier to measure. However, it is a case worth checking, because if there is one possibility that the lines are only seemingly narrow, then this new class of $\gamma$-ray emitting AGN would be reconciled with the common knowledge on blazars and radiogalaxies (and therefore would no more be a new class of $\gamma$-ray AGN). 

There are a few data available on these $\gamma$-NLS1, but one of them -- PMN~J0948$+$0022 -- has been the target of two multiwavelength campaigns in 2009 (Abdo et al. 2009a) and 2010 (Foschini et al. 2011). Therefore, I will study this case as an archetypical example of this new class. 

PMN~J0948$+$0022 ($z=0.585$) was recognized as an anomalous NLS1 by Zhou et al. (2003), who discovered its characteristics of NLS1 together with strong radio emission with a flat spectrum, which in turn suggested the presence of a relativistic jet viewed at small angles (see also Yuan et al. 2008). This specific source was the first of this class to be detected at high-energy $\gamma$ rays by means of \emph{Fermi}/LAT (Abdo et al. 2009b, Foschini et al. 2010).

The optical spectrum was measured by the Sloan Digital Sky Survey (SDSS\footnote{\texttt{http://www.sdss.org/}}) in two different days (2000 February 28 -- MJD 51602; 2000 March 27 -- MJD 51630). Both observations resulted in faint $ugriz$ magnitudes ($\sim 18-19$). The H$\beta$ line has an almost constant $\sigma$: $12.9\pm 0.8$~\AA\, on MJD~51602 and $12.8\pm 0.5$~\AA\, on MJD~51630, while the EW changes slightly, from $16.1\pm 0.8$~\AA \, (MJD~51602) to $21.5\pm 0.7$~\AA \, (MJD~51630). 

During the 2009 MW Campaign, the optical flux dropped in the period 2009 May 5$-$15 and changed also the slope (see Fig.~12 in Abdo et al. 2009a), indicating changes in the synchrotron emission. Given the redshift of $z=0.585$, the H$\beta$ line is at $\sim 7705$~\AA, which is between the $R$ and $I$ filters. Therefore, I calculated the flux densities of these two filters from both the SDSS (2000), from the continuum below the H$\beta$ line and in 2009 early/mid May (MW Campaign). The results are displayed in Table~2.

It is evident that the measurement of the FWHM (Zhou et al. 2003, Yuan et al. 2008) has been done during a period of low continuum flux, i.e. with low jet activity. The period of relatively low jet flux measured during the 2009 MW Campaign is slightly greater and, during the high flux, the values are greater by a factor 3$-$4. As from Sect.~2, the FWHM of the line becomes narrower when the jet continuum increases, but in this case the measurement has been done with the lowest contribution from the jet. Therefore, this is a really NLS1 and the jet activity can only make the line even narrower. 

In addition, it is worth noting that the disc power of NLS1s is much greater than that of BL Lac objects and, hence, the lines are also much more luminous: a situation similar to that of BL Lacs could occur only during exceptional outbursts. The flux density of the line as measured by SDSS is $\sim 1.8\times 10^{-16}$~erg~cm$^{-2}$~s$^{-1}$~\AA$^{-1}$. By comparing this value with $f_{\rm I}$ and $f_{\rm R}$ in the period of high flux (MJD 54956), it is clear that the jet activity literally surpassed the H$\beta$ emission line flux by a factor $\sim 2$. During the 2009 MW Campaign, the source displayed some activity, but is nothing when compared with the more prominent outburst observed in 2010 (Foschini et al. 2011). It is therefore reasonable to think that when the jet is greatly active, the optical spectrum of PMN~J0948$+$0022 could become featureless. This could be verified with a MW campaign, but with optical instruments set to acquire spectra instead of photometry. 

\section{Conclusions}
In this work, I have studied the effect of relativistic jet on the emission line profiles in the cases of BL Lac Objects and $\gamma$-ray Narrow-Line Seyfert 1 Galaxies. In the former case, it is shown that the smallness of the observed FWHM of the Ly$\alpha$ lines, detected by Stocke et (2011) in three BL Lacs, is an effect due to the combined action of both the relativistic jet and a low luminosity accretion disc. Once removed the Doppler boosting of the jet continuum, the intrinsic FWHM of the lines are found to be in the range expected in BLR. 

Instead, in the latter case -- narrow permitted lines in $\gamma$-NLS1s -- it is shown that these lines are really narrow, since the disc and the lines are much more powerful and the measurements of their FWHM were done during periods of low jet activity. Therefore, it is confirmed that $\gamma$-NLS1 is really a new class of $\gamma$-ray emitting AGN, different from blazars and radio galaxies.

\begin{acknowledgements}
I wish to thank the anonymous referee for his/her useful comments, which helped to improve the manuscript.
\end{acknowledgements}

\label{lastpage}


\begin{thebibliography}{99}
\bibitem[Abdo et al. (2009a)]{MW2009} Abdo, A.~A. et al., 2009a, ApJ 707, 727

\bibitem[Abdo et al. (2009b)]{DISCOVERY} Abdo, A.~A. et al., 2009b, ApJ 699, 976

\bibitem[Bachev et al. (2004)]{BACHEV} Bachev, R. et al., 2004, ApJ 617, 171

\bibitem[Blandford \& McKee (1982)]{BLANDFORD} Blandford, R.~D. \& McKee, C.~F., 1982, ApJ 255, 419

\bibitem[Cardelli et al. (1989)]{EXTINCTION} Cardelli, J.~A., Clayton, G.~C. \& Mathis, J.~S., 1989, ApJ 345, 245

\bibitem[Chonis \& Gaskell (2002)]{CHONIS} Chonis, T.~S. \& Gaskell, C.~M., 2008, AJ 135, 264

\bibitem[Collin et al. (2006)]{COLLIN} Collin, S. et al., 2006, A\&A 456, 75

\bibitem[Constantin \& Shields (2003)]{CS} Constantin, A. \& Shields, J.~C., 2003, PASP, 115, 592

\bibitem[Corbett et al. (2000)]{CORBETT} Corbett, E.~A. et al., 2000, MNRAS 311, 485

\bibitem[Elitzur \& Ho (2009)]{ELITZUR} Elitzur, M. \& Ho, L.~C., 2009, ApJ 701, L91

\bibitem[Fine et al. (2011)]{FINE} Fine, S., Jarvis, M.~J. \& Mauch, T., 2011, MNRAS 412, 213

\bibitem[Foschini (2002)]{FOSCHINI} Foschini, L., 2002, A\&A 385, 62

\bibitem[Foschini et al. (2010)]{FOSCHINI1}	Foschini, L. et al., 2010, in: ASP Conference Series vol. 427, Accretion and ejection in AGN: a global view, eds, L. Maraschi, G. Ghisellini, R. Della Ceca \& F. Tavecchio (San Francisco, ASP), 243 [\texttt{arXiv:0908.3313}].

\bibitem[Foschini (2011)]{FOSCHINI3} Foschini, L., 2011, in: Proceedings of Science vol. NLS1, Narrow-Line Seyfert 1 Galaxies and Their Place in the Universe, eds L. Foschini, M. Colpi, L. Gallo, D. Grupe, S. Komossa, K. Leighly, S. Mathur (Trieste, PoS), 024 [\texttt{arXiv:1105.0772}].

\bibitem[Foschini et al. (2011)]{FOSCHINI2} Foschini, L. et al., 2011, MNRAS 413, 1671

\bibitem[Gavignaud et al. (2006)]{GAVIGNAUD} Gavignaud, I. et al., 2006, A\&A, 457, 79

\bibitem[Ghisellini \& Tavecchio (2008)]{NEWSEQUENCE} Ghisellini, G. \& Tavecchio, F., 2008, MNRAS 387, 1669

\bibitem[Ghisellini et al. (2010)]{GHISELLINI} Ghisellini, G. et al., 2010, MNRAS 402, 497

\bibitem[Jarvis \& McLure (2006)]{JARVIS} Jarvis, M.~J. \& McLure, R.~J., 2006, MNRAS 369, 182

\bibitem[Kalberla et al. (2005)]{LAB} Kalberla, P.~M.~W. et al., 2005, A\&A 440, 775

\bibitem[Maoz (2007)]{MAOZ} Maoz, D., 2007, MNRAS 377, 1696

\bibitem[McLure \& Dunlop (2002)]{MCLURE} McLure, R.~J. \& Dunlop, J.~S., 2002, MNRAS 331, 795

\bibitem[Netzer (1990)]{NETZER} Netzer, H., 1990, in: Saas-Fee Advanced Course vol. 20, Active Galactic Nuclei, ed. T. J.-L. Courvoisier \& M. Major (Berlin, Springer), 57

\bibitem[Peterson et al. (1998)]{PETERSON} Peterson, B.~M. et al., 1998, ApJ 501, 82

\bibitem[Pian et al. (2005)]{PIAN} Pian, E., Falomo, R. \& Treves, A., 2005, MNRAS 361, 919

\bibitem[Pian et al. (2010)]{PIAN2} Pian, E. et al., 2010, MNRAS 401, 677

\bibitem[Pounds \& Page (2004)]{OUTFLOWS} Pounds, K. \& Page, K., 2004, Nuclear Physics B (Proc. Suppl.) 132, 107

\bibitem[Rawlings \& Saunders (1991)]{RAWSAU} Rawlings, S., \& Saunders, R., 1991, Nature 349, 138

\bibitem[Shang et al. (2007)]{SHANG} Shang, Z. et al., 2007, AJ, 134, 294 

\bibitem[Stocke et al. (2011)]{STOCKE} Stocke, J.~T., Danforth, C.~W. \& Perlman, E.~S., 2011, ApJ 732, 113

\bibitem[Tavecchio et al. (2010)]{TAVECCHIO} Tavecchio, F. et al., 2010, MNRAS 401, 1570

\bibitem[Telfer et al. (2002)]{TELFER} Telfer, R.~C. et al., 2002, ApJ 565, 773

\bibitem[Vanden Berk et al. (2001)]{VANDENBERK} Vanden Berk, D.~E. et al., 2001, AJ 122, 549

\bibitem[Vestergaard et al. (2000)]{VESTERGAARD} Vestergaard, M., Wilkes, B.~J. \& Barthel, P.~D., 2000, ApJ 538, L103

\bibitem[Wagner (2008)]{WAGNER} Wagner, R.~M., 2008, MNRAS 385, 119

\bibitem[Wandel et al. (1999)]{WANDEL} Wandel, A., Peterson, B.~M. \& Malkan, M.~A., 1999, ApJ 526, 579

\bibitem[Wanders \& Peterson (1996)]{WANDERS} Wanders, I., \& Peterson, B.~M., 1996, ApJ 466, 174 

\bibitem[Wills \& Browne (1986)]{WILLS} Wills, B.~J. \& Browne, I.~W.~A., 1986, ApJ 302, 56

\bibitem[Yuan et al. (2008)]{YUAN} Yuan, W. et al., 2008, ApJ 685, 801

\bibitem[Zhou et al. (2003)]{ZHOU} Zhou, H.-Y. et al., 2003, ApJ 584, 147

\end{thebibliography}
\end{document}